\documentstyle[12pt,epsf]{article}
\newcommand{\baa}{\begin{eqnarray*}}
\newcommand{\eaa}{\end{eqnarray*}}
\newcommand{\bb}{\begin{thebibliography}}
\newcommand{\eb}{\end{thebibliography}}

\newcommand{\lb}{\Lambda}
\newcommand{\lab}[1]{\label{#1}}

\begin{document}
\title{High-energy spin effects and
structure of elastic scattering amplitude  }
%

\author{
Vojt\v{e}ch Kundr\'{a}t, Milo\v{s} Lokaj\'{\i}\v{c}ek\\
{\it
Institute of Physics,
AS~CR, 182 21 Praha 8, Czech Republic}\\
 O.V. Selyugin
\footnote{E-mail: selugin@thsun1.jinr.dubna.su} \\
{\sl Bogoliubov Laboratory of Theoretical Physics,}\\
{\sl Joint Institute for Nuclear Research, Dubna, 141980,
Russia} }

%

\maketitle

\begin{abstract}
A behavior of imaginary and real parts of the  
high-energy elastic hadron scattering amplitude is examined in 
the diffraction region. It is shown that the interference
between Coulomb and hadronic scattering at small momentum 
transfers and especially in the region of diffractive
minimum can bring some information about the structure 
of the hadron spin-non-flip amplitude.

\end{abstract}

\section{Introduction}

One of the goals of modern physics is the exploration of 
internal structures of microscopic matter objects; the 
main attention being devoted to nucleons at present. 
Even if some characteristics of nucleon structure 
may be described on the basis of QCD there are still 
many open questions, as the most experimental data 
on diffractive scattering have been interpreted with 
the help of phenomenological models. The theory has 
concerned mainly the nucleon structure playing a role 
in deep inelastic nucleon scattering while a similar 
theory of diffractive collisions has not yet been
practically available. 

In the most phenomenological approaches to
high-energy elastic hadron scattering one uses
the elastic hadron scattering amplitude with 
a dominant imaginary part vanishing at the 
diffractive minimum \cite{bloc,matt}; its 
real part is introduced only to fill 
in the non-zero value of differential cross section
${d \sigma / dt}$ in the diffraction dip. The 
corresponding elastic phase remains very small
in a rather broad interval of momentum transfers 
$t$. Performing Fourier-Bessel transformation
of such an amplitude one always obtains a central behavior
of elastic hadron scattering in the impact parameter space,
i.e., with the majority of elastic hadronic collisions 
realized at impact parameter value $b = 0$; colliding 
nucleons are interpreted as transparent objects \cite{miet}.

However, the existence of diffractive minimum does not
mean at all that the imaginary part of the hadronic amplitude
should vanish in it; it means only that the sum of the squares
of both its real and imaginary parts has a minimum. It 
has been shown \cite{kun3} that experimental data 
may be well described with the help of amplitude 
giving a peripheral behavior (having maximum at 
a higher value of $b$). In such a case
the imaginary part of hadronic amplitude 
is dominant in a much smaller region of $t$ only.
The peripheral interpretation of elastic hadron 
scattering at high-energy enables a unified 
description of all diffractive processes, 
i.e., of both the elastic and diffractive 
production channels. 
 
The behavior of high-energy elastic hadron collisions in
the impact parameter space can be simply characterized 
by the values of root-mean-squares of impact parameters
corresponding to the elastic, inelastic and as well 
as total scattering which can be easily calculated
directly from the $t$ dependent elastic hadron scattering 
amplitude \cite{kune}. These quantities characterize the ranges 
of forces responsible for all mentioned kinds of scattering.
The values of elastic and inelastic root-mean-squares of
impact parameters differ significantly in the case of 
the central and peripheral interpretations of elastic 
hadron scattering.

As the interference region of the Coulomb and hadronic
interaction is very narrow the elastic scattering 
data themselwes do not allow to decide between mentioned
different alternatives. And it is necessary to look for
other kinds of experiments. One possibility may consist
in experiments with polarized nucleons.
Such experiments are planned to be performed 
at RHIC and LHC \cite{akpr,gur,nur}. They include 
the measurement of spin correlation parameters 
in the diffractive region of $pp$ elastic scattering. 
The goal of this paper is to investigate whether they
might contribute to giving an answer to the given
question. The individual spin components exhibit
different $t$ dependence for peripheral or central
behaviors which may manifest in different predictions
concerning the analyzing power at higher $|t|$ values.

\section{Elastic hadron scattering amplitudes}

 The differential cross sections 
determined in an experiment are described by the absolute square 
of the total scattering amplitude including the electromagnetic
and hadronic forces
\begin{eqnarray}
  F(s,t) =
  F_{C}(s,t) \exp{(i \alpha \varphi (s,t))} + F_{N}(s,t),
\label{ta}
\end{eqnarray}
where $F_{C}(s,t)$ stands for the Coulomb amplitude known 
from QED and $F_N(s,t)$ is the elastic hadronic amplitude;
$\alpha = 1/137$ is the fine structure constant.
For the relative phase $\varphi (s,t)$ between the Coulomb
and hadronic amplitudes various expressions are used.
In the usual approaches the West and Yennie simplified
formula \cite{west} valid for very small values of $|t|$ is used.
However, for the analysis at higher values of $|t|$ more
precise formula seems to be preferable \cite{selprd,selmpl,selpl}
\begin{eqnarray}
  \varphi(s,t) =  \varphi_{C}(s,t) - \varphi_{CN}(s,t),
\end{eqnarray}
where $\varphi_{C}(s,t)$ stands for the second Born approximation of
the pure Coulomb amplitude and the term $\varphi_{CN}(s,t)$ is
defined by the Coulomb-hadron interference. It has been found that
the interference of the hadronic and electromagnetic amplitudes
may give an important contribution not only at very 
small transfer momenta but also in the region of the
diffraction minimum \cite{selprd}. For this purpose 
one should know the $\varphi (s,t)$
at sufficiently large momentum transfers, too.

 As a standard amplitude let us take the amplitudes proposed
in Ref. \cite{zpc}. It has been shown on the basis of sum rules
\cite{selyf} that the main contribution to
hadron interaction at large  distances  comes  from  the  triangle
diagram with $2\pi $ meson exchange in the $t$ channel. As a result, the
hadron amplitude can be  represented  as  a  sum  of  central  and
peripheral parts of the interaction
\begin{eqnarray}
F_N(s,t) \propto T_{c}(s,t) + T_{p}(s,t),
\end{eqnarray}
where $T_{c}(s,t)$ describes the
interaction between the central parts of hadrons; 
and $T_{p}(s,t)$ is the sum of contributions of 
diagrams corresponding to the interactions 
of the central part of one hadron with the meson
cloud of the other. The contribution of these diagrams
to the scattering amplitude  with the $N(\Delta $ isobar) in the
intermediate state looks as follows \cite{zpc}:
\begin{eqnarray}
T ^{\lambda _{1}\lambda _{2}}_{N(\Delta )}(s,t) =
{g^{2}_{\pi NN(\Delta )}\over i(2\pi )^{4}}\int
 d^{4}q T_{\pi N}(s\acute{,}t)\zeta _{N(\Delta )}
[(k-q),q^{2}]\zeta_{N(\Delta )}[(p-q),q^{2}]  \nonumber \\
\times \frac{\Gamma^{\lambda_{1}\lambda_{2}}(q,p,k,)}
{[q^{2}- M^{2}_{N(\Delta )}+ i\epsilon ][(k-q)^{2}- \mu ^{2}+i\epsilon ]
[(p-q)^{2}- \mu^{2}+ i\epsilon ]}.
\end{eqnarray}
Here $\lambda_{1},\lambda_{2}$ are helicities of nucleons,
$g^{2}_{\pi NN(\Delta )}$ is the coupling constant,
$T_{\pi N}$ is the
$\pi N$ scattering amplitude, $\Gamma$ is the matrix element of the numerator
of the representation of the diagram and $\zeta$ are vertex functions chosen
in the dipole form with the parameters $\beta _{N(\Delta )}$:
\begin{eqnarray}
\zeta_{N(\Delta )}(l^{2},q^{2}\propto  
M^{2}_{N(\Delta )}) = {\beta ^{4}_{N(\Delta )}
\over (\beta ^{2}_{N(\Delta )}- l^{2})^{2}}.
\end{eqnarray}
   The model with the $N$ and $\Delta$
contributions provides a self-consistent 
picture of the differential cross section
of different hadron processes at high energies.
Really, parameters in the  amplitude determined from one
reaction, for example, elastic $pp$ scattering, allow one 
to obtain a wide range of results for elastic
meson-nucleon scattering and charge-exchange reaction
$\pi^{-} p \rightarrow  \pi^{0} n$ at high energies.

The preceeding approach of separating the Coulomb 
and hadronic scattering is convenient in the
case of usual weakly $t$ dependent phase giving
the central description of elastic hadron collisions
in the impact parameter space. However, when the
$t$ dependent phase strongly increases with the
increasing $|t|$ the more convenient method
of separation of Coulomb and hadronic scattering
consists in the application of the approach proposed
in Refs. \cite{kun3,kun4} valid at any $s$ and $t$
(up to the terms linear in $\alpha$). However, the dynamics
of hadronic amplitude leading to peripherality has not 
been up to now theoretically specified. In principle
it could be described in different ways, e.g., in the simple 
Regge pole model with the help of complex $t$ dependent 
residue functions, complex Regge trajectories, etc.
Instead of it, in an analogy with the use of
current analysis based on the West and Yennie 
amplitude \cite{west}
a convenient parameterization of both the modulus
and the phase has been used (for detail, see Ref.
\cite{kun3}). Such an approach enables to specify
the elastic hadron scattering amplitude $F_N(s,t)$
directly from the elastic scattering data.

The difference between the phases leading either to
central or peripheral distributions of elastic hadron scattering
in the impact parameter space can be seen in Fig. 1;
the graphs correspond to $\bar{p}p$ scattering at energy of 
541 GeV \cite{kun3}; similar $t$ dependencies may be also
exhibited by the phases corresponding to the $pp$ elastic 
scattering at energy of 53 GeV (see Ref. \cite{kun3}).

\vspace{2.cm}

\begin{center}
\hspace*{1.8cm}
\epsfysize=7.cm
\epsfxsize=15.cm
\epsfbox{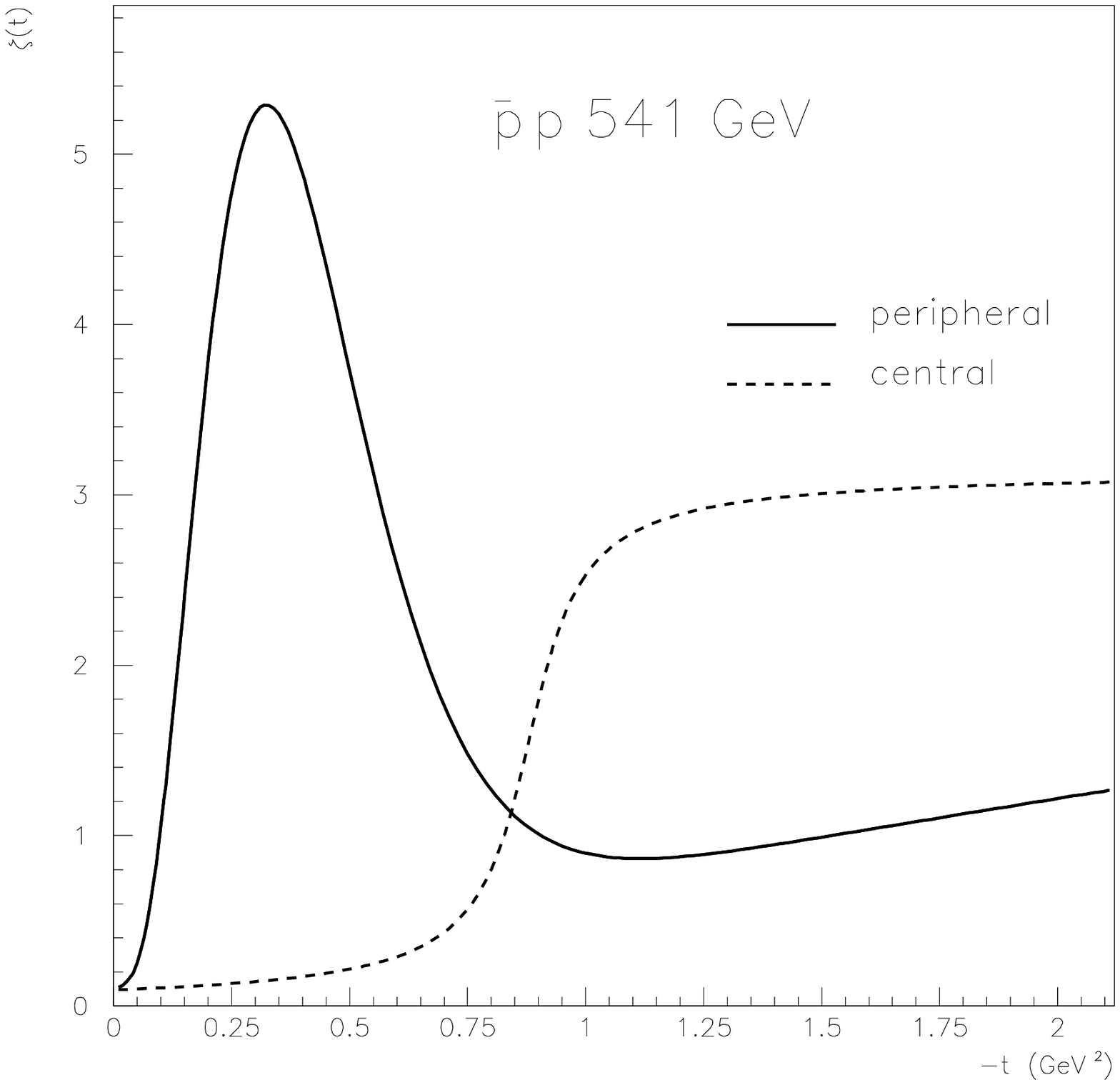}
\end{center}
\vspace*{-2cm}
Fig. 1.
The $t$ dependent phases of elastic hadron scattering amplitudes
leading to peripheral (solid line) and central (dashed line)
distributions of elastic hadron scattering in the impact parameter 
space.

\section{Analysis of $A_N$}

Now, let us examine the form of analyzing power 
of $pp$  elastic scattering using the previous 
specifications of the scattering amplitudes.
The differential cross section ${d \sigma \over dt}$
and spin parameters $A_N$ are defined as
 \begin{eqnarray}
  \frac{d\sigma}{dt}&=& \frac{2
\pi}{s^2}\;\;\;(|\Phi_1|^2+|\Phi_2|^2+|\Phi_3|^2
   +|\Phi_4|^2+4|\Phi_5|^2),
\lab{dsdt}
\\
  A_N\frac{d\sigma}{dt}&=& -\frac{4\pi}{s^2}\;\;\;
                 \Im[(\Phi_1+\Phi_2+\Phi_3-\Phi_4) \Phi_5^{*})],
\lab{an}
\end{eqnarray}
where $\Phi_j$ are usual helicity amplitudes.
It has been shown in Refs. \cite{bttled} that due to
presence of the Coulomb and hadronic interactions
each total helicity amplitude $\Phi_j(s,t)$
can be written as the sum of the hadronic 
$\Phi_j^N(s,t)$ and electromagnetic $\Phi_j^{elm}(s,t)$
helicity amplitudes bound with the same relative phase
$\varphi(s,t)$ valid for small $|t|$ values
\begin{equation}
\Phi_j(s,t) = \Phi_j^N(s,t) + \Phi_j^{elm}(s,t) e^{\varphi(s,t)},
\end{equation}
where the electromagnetic helicity amplitudes can be
explicitly determined within the framework of QED.
Their limiting values at high energies have been
given in Ref. \cite{selprd}; only three of them are
independent at asymptotic energies.

At present, the spin effects owing to the 
Coulomb-nucleon interference (CNI) at very
small transfer momenta have been widely discussed 
in the relation to future spin experiments 
at RHIC and LHC. These effects had not been well
understood in the domain of the diffraction 
minimum. This is due to the fact that the
Coulomb-hadron interference phase at higher
$|t|$ has not been known sufficiently. 
In \cite{selmpl}, the phase $\varphi_c$ of 
the pure Coulomb amplitude in the second 
Born approximation with the form factor 
in the monopole and dipole forms has been 
calculated in a broad region of $t$. It has 
been shown that the behavior of $\varphi_c$ 
at higher values of $t$ sharply differs from the 
behavior of $\varphi_c$ obtained in \cite{can}.

For the total phase factor that can be used in 
the whole diffraction region of elastic hadron 
scattering it has been found \cite{selprd}
\begin{eqnarray}
 \varphi(s,t) = \ln{\frac{q^2}{4}} +2\gamma +\frac{1}{F_N(s,q)}
  \int_{0}^{\infty} \tilde{\chi}_{c}(\rho)
  (1 - \exp(\chi_h(\rho,s))J_{0}(\rho,q)d\rho , \lab{fei2}
\end{eqnarray}
with
\begin{eqnarray}
  \tilde{\chi}_c(\rho) = 2\rho \ln{\rho} +2\rho K_{0}(\rho \lb)
  [1+ \frac{5}{24} \lb^2 \rho^2 ]
   +\frac{\lb \rho}{12} K_1(\rho \lb) [11+ \frac{5}{4} \lb^2 \rho^2].
\end{eqnarray}
Here $K_0$ and $K_1$ are modified Bessel functions of the zero
and first orders, $\lb = 0.71 $GeV$^2$, $\gamma = 0.577$ is Euler
constant and $t = -q^2$.
Then the contributions of the Coulomb-hadron interference
to the analyzing power $A_N^{CN}$ (including the 
two possible structures of spin-non-flip amplitude 
at small transfer momenta and in the diffraction 
dip domain) of the $pp$ elastic scattering
at small $|t|$ for $\sqrt{s} = 540$ GeV is
shown in Fig. 2.

The obtained form of $A_{N}^{CN}$ at small momentum transfers
differs for the two variants beginning at $|t| > 0.05$ GeV$^2$.
The difference reach $2 \%$ at $-t = 0.15 \ GeV^2$ and,
in principle, can be measured in accurate experiment.

Now, let us calculate the Coulomb-hadron interference effect
in the analyzing power $A_{N}^{CN}$
in the two alternatives for higher $|t|$: \\
   (i) the diffraction dip is created by the "zero" of the imaginary
       part of the scattering amplitude and the real part fills it in \\
   (ii) the diffraction dip is created by the "zero" of the real part
       part of the scattering amplitude and the imaginary part fills it in \\

  The results are shown in Fig. 3 for $\sqrt{s} = 50$ GeV and
  in Fig. 4 at $\sqrt{s}= 540$  GeV.
  
\epsfysize=7.cm
\epsfxsize=10.cm
\centerline{\epsfbox{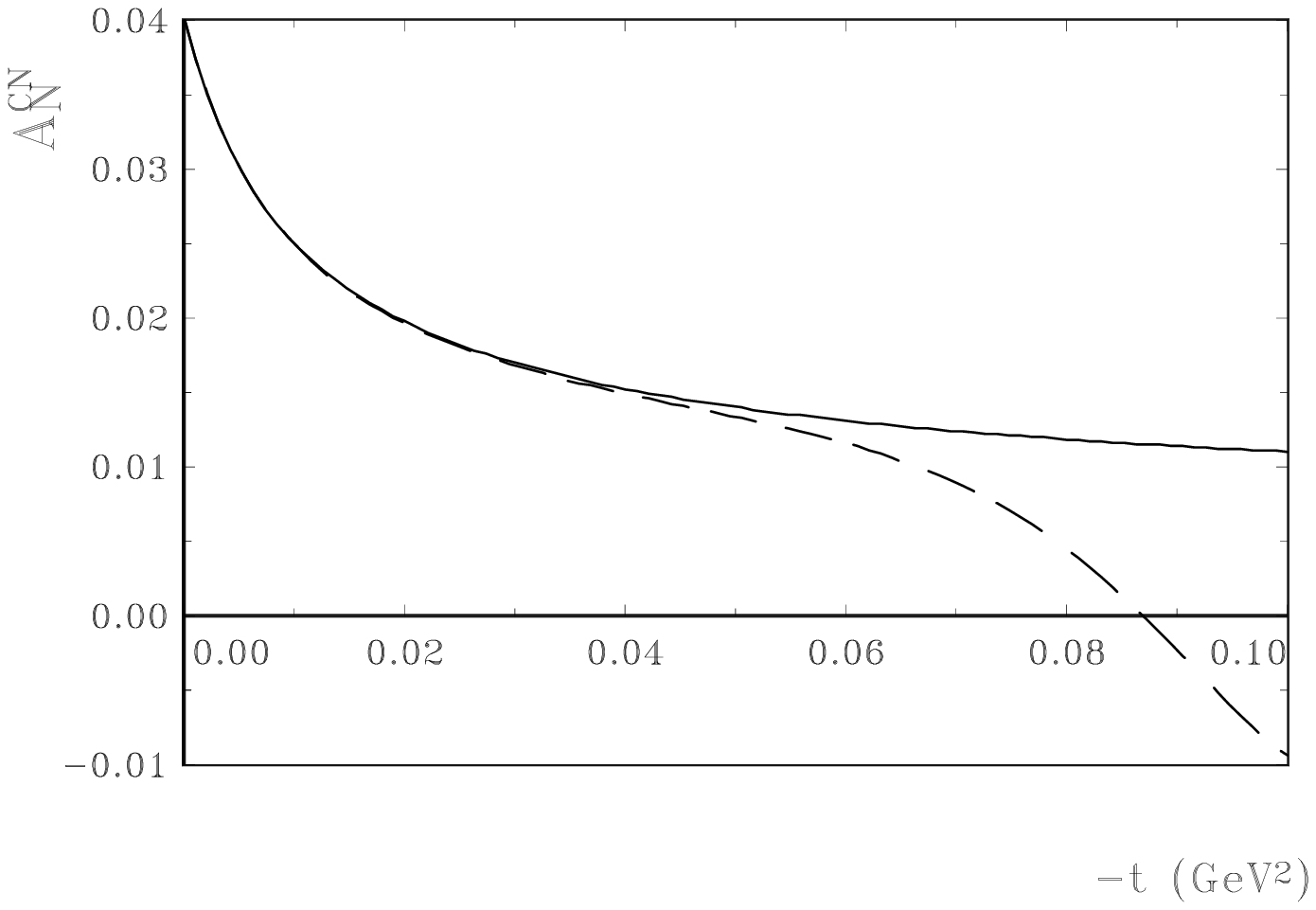}}
Fig. 2.
 Calculated analyzing power (the solid line) 
 corresponds to the first variant characterized by
 zero of the real part at small $|t|$; the dashed  
 line corresponds to the second variant with 
 imaginary part vanishing at small $|t|$.
 
\epsfysize=7.cm
\epsfxsize=10.cm
\centerline{\epsfbox{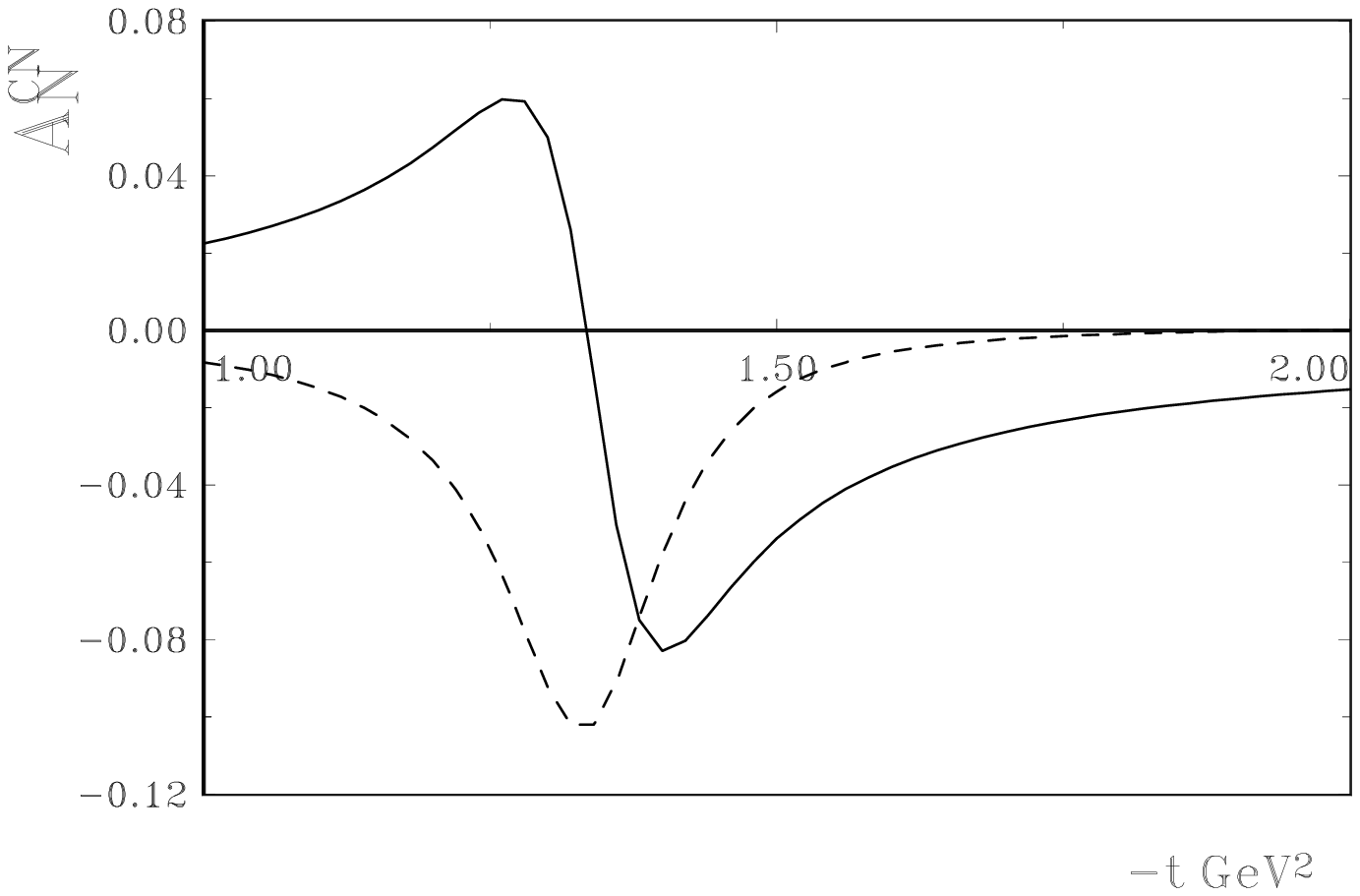}}

Fig. 3.
 Calculated analyzing power in the region of the diffraction minimum
  at $\sqrt{s}=50 $ GeV
  (the solid line corresponds to the variant I with zero of $\Im F_{N}$
   at dip; the dashed line shows the variant II, with the
   zero of the $\Re F_{N}$ at the dip).

\epsfysize=7.cm
\epsfxsize=10.cm
\centerline{\epsfbox{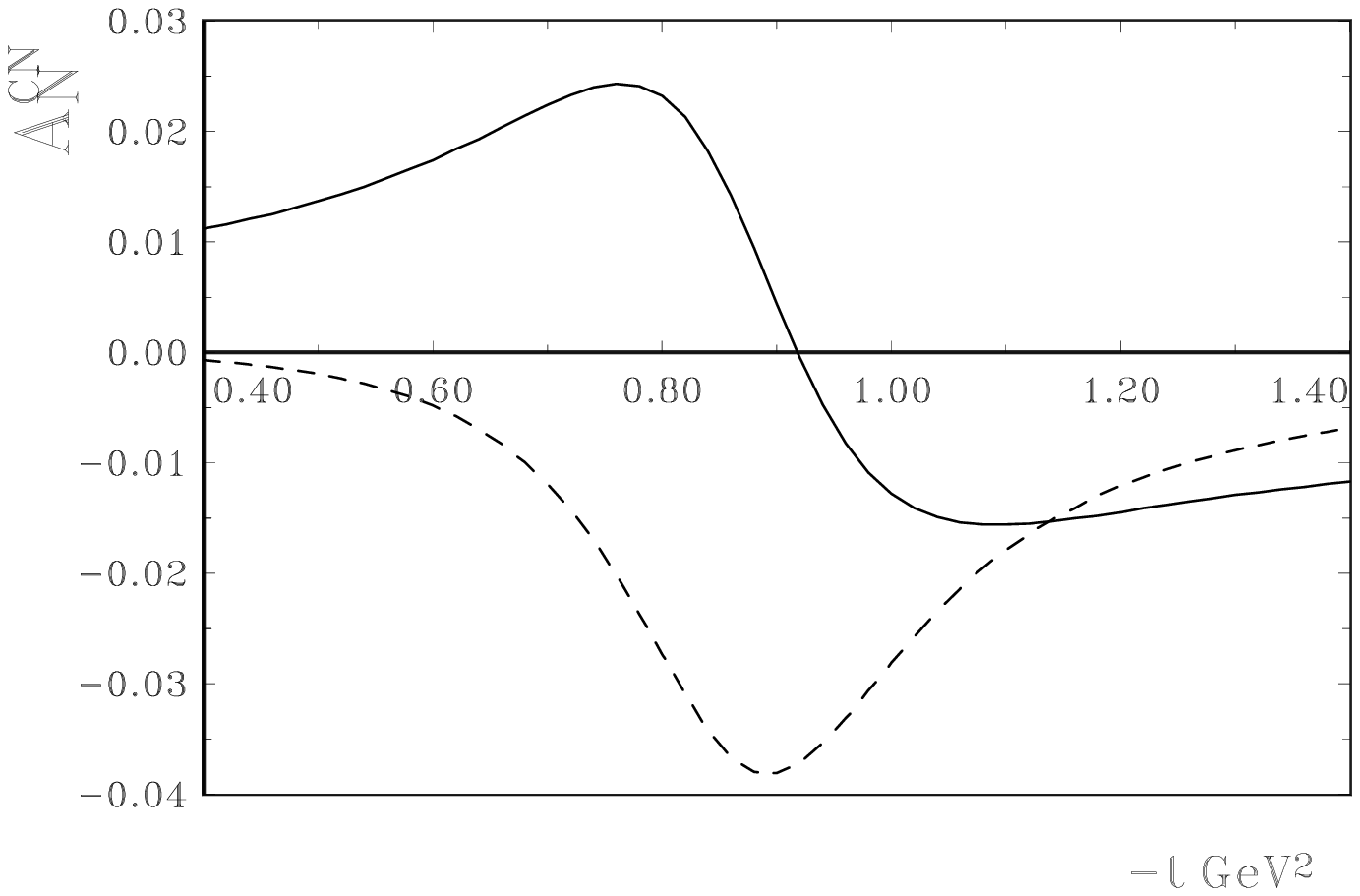}}
Fig. 4.
 Calculated analyzing power in the region of the diffraction minimum
  at $\sqrt{s}=540 $ GeV
    (the solid line corresponds to the variant I with zero of $\Im F_{N}$
   at dip; the dashed line shows the variant II, with the
   zero of the $\Re F_{N}$ at the dip).

\vspace*{1cm}
In the region below the diffraction minimum, we obtain in 
the first case a positive non-small contribution which
changes (reduces) the size of the spin correlation
parameter owing to the hadron-spin-flip amplitude.
Especially it is to be noted that there is a heavy
shift of the point where   
$A_N$ changes its sign. The calculations show
that such additional contributions lead to the
change of the position of the negative maximum 
of $A_N$ to larger $|t|$. In the other case, 
with the zero of the real part of the scattering
amplitude, we obtain only negative contribution 
which enhanced the magnitude of $A_N$ owing to the 
hadron-spin-flip amplitude and change the position 
of maximum to the low $|t|$. Such a picture corresponds to 
the highest energy at RHIC. Of course, $A_N^{CN}$ has 
small magnitude but can be measured in an
accurate experiment.

\section{Conclusion}

We have shown that the two different $t$ dependent
phases of elastic hadron scattering amplitude 
giving the central and peripheral distributions of
elastic hadron scattering in the impact parameter 
space can give also different predictions of 
the spin correlation parameter - the analyzing power
$A_N$ at higher values of $|t|$; mainly in the region 
of diffractive minimum. This is due to the the interference 
of the spin-non-flip elastic scattering amplitude and the
electromagnetic spin-flip amplitude.
The measurement of the correction $A_{N}^{CN}$ in the domain
of the diffraction minimum can give valuable information 
about the structure of the elastic scattering amplitude.
   In spite of the large  contribution of the hadron-spin-flip
amplitude, we can see that taking into account the odderon
contributions leads to visible changes in spin correlation effects.
So, the precise measurement of $A_N$ in the region
of the diffraction minimum and the treatment of their energy
dependence can give some additional information which would allow 
to define the sign and magnitude of the odderon contribution, too.
One should mention the importance of the $t$ dependence of the 
odderon amplitude, and therefore, the possible bounds on 
the $t$ dependence of the odderon amplitude might be also 
revealed in the given experiments.

\end{document}